# A Parametric Framework for Anticipatory Flashflood Warning: Integrating Landscape Vulnerability with Precipitation Forecasts

Xiangpeng Li[1], Junwei Ma[1*], Samuel D Brody[2], Ali Mostafavi[3]

[1] Ph.D. student. Urban Resilience.AI Lab, Zachry Department of Civil and Environmental Engineering, Texas A&M University, College Station, Texas, United States.

[2] Director, Institute for a Disaster Resilient Texas (IDRT)Regents Professor, Marine and Coastal Environmental Science, College of Marine Sciences & Maritime Studies, Texas A&M University at Galveston

[3] Professor. Urban Resilience.AI Lab, Zachry Department of Civil and Environmental Engineering, Texas A&M University, College Station, Texas, United States.

[*] Corresponding author: Junwei Ma, E-mail: jwma@tamu.edu.

**Abstract**

Flash flood warnings are largely reactive, providing limited advance notice for evacuation planning and resource prepositioning. This study presents and validates an anticipatory, parametric framework that converts landscape vulnerability and precipitation into transparent, zone-aware threat levels at neighborhood scales. We first derive an inherent hazard likelihood (IHL) surface using pluvial flood depth, height above nearest drainage, and distance to streams. Next, we compute a hazard severity index (HSI) by normalizing 24-hour rainfall against local Atlas-14 100-year, 24-hour depths. We then integrate IHL and HSI within a localized threat severity (LTS) matrix using 20 class-specific triggers, requiring lower exceedance in high-risk terrain and higher exceedance in uplands. Applied to two Texas flood events, the LTS exhibits statistically significant spatial association with independent crowdsourced impact proxies, capturing observed disruption hotspots. The framework is computationally lightweight, scalable, and extends actionable situational awareness into a 48-72 hour anticipatory window, supporting pre-event decision-making by emergency managers.

## 1. Introduction

Flash flood risk is rising with urbanization and climate variability, yet most operational systems still rely on reactive hydrologic-hydraulic workflows that issue alerts only as inundation develops[1]. While these methods deliver precision at short lead times, they leave a critical 24- to 72-hour anticipatory window underserved and often out of reach for resource-constrained jurisdictions due to data and computational demands[2,3]. Simultaneously, widely available terrain-and exposure-based indicators, such as height above nearest drainage (HAND), distance-to-stream, and national pluvial hazard products, are predominantly used for static susceptibility mapping rather than for forecast-conditioned, operational warning[3,4]. This disconnect between long-term risk layers and real-time alerts motivates an alternative approach: a parametric, design-storm-normalized framework that fuses baseline landscape vulnerability with event severity, producing warnings that are transparent, auditable, and deployable at neighborhood and watershed scales before flooding begins. Also, flash flood warning science and practice remain largely reactive (e.g., issuing alerts only as inundation develops), leaving a poorly served 24- to 72-hour anticipatory window when the most critical logistical decisions must be made, including evacuation staging and inter-agency coordination. This window is particularly problematic for jurisdictions that lack the data, computational resources, and expertise to run detailed hydrologic-hydraulic models[5]. This gap motivates our objective to operationalize a parametric, forecast-conditioned framework that transforms widely available static landscape indicators and precipitation-frequency benchmarks into transparent, zone-aware triggers before an event. Moreover, the current approaches often require data, computational resources, and expertise that many jurisdictions cannot sustain. At the same time, while terrain-based susceptibility metrics (e.g., HAND, distance-to-stream) and national flood products (e.g., Fathom) are well-established individually, they are rarely operationally fused with climatological precipitation-frequency benchmarks (e.g., National Oceanic and Atmospheric Administration (NOAA) Atlas 14) to yield zone-aware, parametric thresholds that can be applied days in advance of an event. This gap between long-term risk maps and real-time, sensor-driven warnings leaves an underserved anticipatory window where decisions with the greatest logistical footprint must be made. Because prevailing flash flood systems are largely reactive and resource-intensive, they limit anticipatory action in the 24- to 72-hour window.

The objective of this study is to design and validate a parametric, forecast-conditioned framework that operationalizes early, zone-aware warnings using open national datasets. We first compute an Inherent Hazard Likelihood surface from Fathom pluvial depth, HAND, and distance-to-stream to encode baseline susceptibility. Second, we derive 24-hour radar rainfall as a Hazard Severity Index through Multi-Radar Multi-Sensor (MRMS)-to-Atlas 14 ratios that preserve local climatology. Third, we fuse these components on an H3-L10 grid into a Localized Threat Severity matrix with class-specific thresholds, then empirically test whether higher LTS levels align with independent impact proxies, specifically 311 service requests and Waze traffic/incidents, for two contrasting Texas floods: Tropical Storm Imelda in 2019 and the Dallas flash flood in 2022. The goal of this analysis is to produce an auditable, computationally lightweight framework that extends actionable situational awareness days ahead while complementing real-time hydrologic-hydraulic models in operational practice.

This work advances flood forecasting and management by formalizing an anticipatory, parametric framework that fuses static landscape vulnerability with event-specific precipitation severity into a single, auditable signal at operational scale. Methodologically, we (i) derive IHL from HAND, stream proximity, and pluvial hazard to encode baseline susceptibility; (ii) transform forecast and observed 24-hour rainfall into a HSI using NOAA Atlas 14 precipitation-frequency ratios, thereby preserving local climatology; and (iii) couple all data in an H3-grid LTS matrix with class-specific thresholds. This design yields transparent, zone-aware triggers (expressed as percent-of-design storm), computationally lightweight, and offer prospectively testable days, representing a clear departure from reactive, sensor-driven warnings. Validation against independent, crowdsourced impact proxies demonstrates that the LTS signal captures spatial gradients of consequence across contrasting storm regimes, thereby establishing a generalizable foundation for anticipatory flash-flood impact assessment. This framework enables timely localized decisions for resource prepositioning, targeted messaging via conventional and social media, and road closure recommendations from tiered LTS outputs at neighborhood and watershed granularity. The same parametric index underpins pre-arranged financing for rapid-response budgets. Its cloud-ready data pathways (MRMS/radar, Atlas 14, open topography) and simple audit trail make the framework readily deployable by resource-constrained agencies.

## 2. Background

Flood risk modeling has evolved substantially in recent decades[6], reflecting advances in computational hydrology[7,8], remote sensing[9], and data analytics[10,11] aimed at mitigating the rising impacts of floods intensified by climate change and rapid urbanization. This evolution of flood risk modeling is marked by a shift from traditional hydraulic and hydrologic approaches to more integrated and multidisciplinary methods that incorporate urban planning and land use management[12]. Operational flash flood warning systems have traditionally centered on real-time monitoring and hydrologic-hydraulic (H&H) modeling approaches[13–16]. These systems leverage precipitation radar[17] and physics-based hydrodynamic models[18,19] to generate alerts as flooding conditions develop, typically providing real-time forecasting[20,21]. These traditional one-dimensional (1-D) and two-dimensional (2-D) hydraulic models have long been the foundation of flood hazard mapping and risk assessment[22,23]. Recent studies emphasize the importance of hybrid modeling approaches that combine physical and data-driven methods, leveraging machine learning and deep learning techniques to enhance prediction accuracy and response systems[24]. The integration of internet of things (IoT) sensor networks and remote sensing technologies has become indispensable for real-time flood monitoring and prediction, offering high-resolution data that improve model precision and adaptability[25,26]. The development of large-scale hydrological models and the use of advanced technologies like data assimilation and adaptive grid resolutions have further refined flood prediction capabilities, supporting sustainable watershed management strategies[26,27]. Additionally, the adoption of frameworks like the integrated risk linkages (IRL)

Framework provides a systematic approach to flood risk assessment, arising from the intersection of hazard, vulnerability, and resilience[28]. As urban areas continue to grow, the susceptibility to flooding increases, necessitating the development of more effective urban flood management practices that incorporate remote sensing and machine learning for susceptibility mapping[29]. The HAND mode has emerged as an effective terrain-based indicator for identifying flood-prone areas using only digital elevation models (DEMs) and drainage networks[30–35]. HAND estimates each location's vertical distance from the nearest drainage line, enabling quick delineation of low-lying areas susceptible to inundation[36–39]. Numerous studies have demonstrated HAND's capability to reproduce inundation patterns comparable to those generated by hydrodynamic simulations, especially where detailed hydraulic parameters are unavailable[22,23]. Areas in close proximity to rivers experience enhanced exposure to multiple flood mechanisms, including overbank flooding, backwater effects, and rapid water table rise from stream-aquifer interactions.[40–42] Consequently, proximity to rivers has been identified as a primary explanatory variable in flood susceptibility models, consistently demonstrating strong predictive power in delineating riverine flood hazard zones[43–45]. Parallel to these terrain-driven methods, physically based national models have expanded flood mapping to unprecedented spatial scales. Fathom's flood risk data is a critical resource for understanding and mitigating the impacts of flooding across various regions. The Fathom Global Flood Map serves as a comprehensive tool that provides insights into multiple flood perils, including pluvial, fluvial, and coastal flooding, which are essential for effective risk assessment and management. For example, Gold and Steinberg-McElroy (20)[46] calculate these estimates by multiplying the ratio of residential building footprint area intersecting high flood hazard areas with 2020 decennial census block counts. This dataset enhances flood resilience planning by offering updated and detailed information on populations at risk of flooding in the contiguous US. The ongoing research and collaboration across disciplines are essential for advancing flood modeling techniques and developing adaptive management strategies to mitigate the impacts of climate change on urban flooding[25,47]. These efforts highlight the critical need for innovative approaches to improve prediction, preparedness, and response mechanisms in flood risk management[48]. While effective for immediate emergency response during active events, this reactive framework offers limited opportunity for anticipatory actions that require extended preparation time, such as evacuation planning, resource prepositioning, and community mobilization. Additionally, challenges such as data scarcity, high computational demands, and the need for improved uncertainty quantification persist[24,26]. Furthermore, the implementation of detailed flood models requires substantial computational resources, technical expertise, and comprehensive data infrastructure—barriers that many resource-constrained communities face.

Recognizing the need for complementary approaches that extend warning horizons while remaining operationally scalable, this study develops a parametric framework for anticipatory flash flood warning. The framework integrates pre-existing landscape vulnerability with forecasted precipitation extremes to identify high-risk areas days in advance of flood events, enabling proactive decision-making without reliance on computationally intensive hydrodynamic simulations. Specifically, it combines three parametric indicators: (1) IHL, derived from Fathom pluvial flood depth, height above nearest drainage, and distance-to-stream metrics to represent baseline landscape susceptibility; (2) HSI, computed from ratios of forecasted MRMS rainfall to Atlas 14 design-storm depths to quantify event-specific intensity; and (3) These components are synthesized to generate localized threat severity classifications at the H3 hexagonal grid level, producing spatially explicit flood threat assessments. The framework is applied to two major flood events (i.e., Hurricane Imelda (2019) in Harris County and the August 2022 Dallas flood) to evaluate performance across distinct hydrologic settings. Importantly, this approach is not intended to replace existing real-time warning systems, but to complement them by extending the temporal window of actionable insights. By providing early situational awareness several days before a flood, the parametric framework bridges the gap between long-term flood risk assessment and real-time emergency response, offering a scalable and accessible tool for anticipatory disaster preparedness.

## 3. Study areas and datasets
### 3.1 Study areas

This study focuses on two flood-prone urban counties in Texas (Harris County and Dallas County) as representative case studies. Harris County, which encompasses the Houston metropolitan area, is characterized by low-lying coastal plains with an extensive bayou network and extensive impervious surface coverage[49]. The county experiences frequent flooding events driven by tropical cyclones and convective rainfall systems. Tropical Storm Imelda, which serves as the representative event for this study, delivered more than 430 mm of rainfall within 48 hours (September 17–19, 2019) and causing widespread inundation and infrastructure disruptions across the county[50].

Dallas County, located in north-central Texas within the upper Trinity River basin exhibits a mixture of urbanized areas and rolling uplands. Flooding in Dallas County is predominantly driven by short-duration, high-intensity convective storms[51]. The August 2022 flash flood event provides an illustrative case, resulting from localized rainfall that exceeded the 100-year design threshold, particularly in the eastern part of the county[52].

Examining these two cases allows evaluation of the framework's transferability between coastal-tropical and inland-continental rainfall regimes, as well as between flat alluvial and upland fluvial geomorphic settings.

### 3.2 Data description

The analytical framework integrates three categories of datasets: (1) static geophysical indicators of baseline flood susceptibility, (2) dynamic precipitation observations defining event severity, and (3) crowdsourced impact datasets for empirical validation. All spatial datasets were aggregated to the H3 hierarchical hexagonal grid system for comparative analyses. A summary of all datasets used in this study is presented in Table 1.

### 3.2.1 Static flood susceptibility indicators

- **Fathom Pluvial Flood Depths**

Flood hazard characterization in this study utilizes Cursory Floodplain Data 2025, a statewide dataset developed by Fathom for the Texas Water Development Board[53]. The dataset provides high-resolution flood-hazard layers across Texas, covering fluvial (riverine), pluvial (surface water), and coastal (storm surge) flooding for multiple return periods. This flood-hazard layer captures flooding due to extreme rainfall overwhelming local drainage systems, a hazard particularly relevant in urban contexts where surface flooding contributes significantly to disruption and damage. The Fathom framework is based on the LISFLOOD-FP 1D–2D hydrodynamic model, which simulates water movement across river channels, floodplains, and urban surfaces[54]. The data is available for five annual exceedance probabilities (AEPs): 20% (1-in-5), 10% (1-in-10), 4% (1-in-25), 1% (1-in-100), and 0.2% (1-in-500). Each peril is distributed as 3-meter resolution raster tiles and polygon flood extents. In this study, we employ the pluvial 1-in-100-year (1%) flood hazard scenario.

- **Distance to Stream**

Two stream proximity layers were derived from the U.S. Geological Survey (USGS) National Hydrography Dataset (NHD) to quantify hydrologic connectivity within the study areas[55]. Euclidean distances were computed from each raster cell to the nearest stream polyline using GIS-based spatial analysis. The first layer, distance-to-stream$_0$, captures proximity to all watercourses, including tertiary drains and culverts. The second layer, distance-to-stream$_4$, isolates primary channel systems (Strahler stream order ≥4), typically with contributing areas exceeding 50 km². These two layers capture differential hydrologic connectivity between minor and major drainage systems, with closer proximity generally corresponding to higher flood likelihood. Both variables were later standardized to a 0–1 scale by min–max scaling for multi-criteria integration.

- **Height Above Nearest Drainage**

The height above nearest drainage metric quantifies the vertical distance between a given location and its nearest drainage channel, providing a terrain-based indicator of relative flood susceptibility[56]. Areas with lower HAND values are situated closer to drainage features and are therefore more prone to inundation during extreme rainfall or other flood-inducing events. In this study, HAND serves as a topographic proxy for potential flood exposure, reflecting the degree to which local elevation influences hydrologic flow accumulation. The HAND dataset was obtained from the University of Texas National Flood Interoperability Experiment continental flood inundation mapping system, provided at a 10-meter spatial resolution[57]. For analysis, HAND values were aggregated by computing the mean HAND across all raster pixels, producing a representative measure of average elevation above the nearest drainage within each spatial unit.

### 3.2.2 Dynamic rainfall severity metrics

- **Multi-Radar Multi-Sensor 24-hour accumulated rainfall**

The Multi-Radar Multi-Sensor (MRMS) system, developed by the NOAA National Severe Storms Laboratory, provides radar-based quantitative precipitation estimates at 1-km spatial and 5-minute temporal resolution[58]. The system integrates multiple data sources, including dual-polarization NEXRAD radar, rain gauge observations, and satellite-derived precipitation products, through a quality-controlled data fusion framework to generate spatially continuous rainfall fields.

For this study, 24-hour accumulated rainfall totals were extracted from the MRMS dataset corresponding to the primary flood impact periods for each case event. In Harris County, rainfall data were obtained for Tropical Storm Imelda, which occurred from September 17–19, 2019, while in Dallas County, rainfall accumulation was analyzed for the August 2022 flash flood event, spanning August 20–23, 2022.

- **NOAA Atlas 14 precipitation frequency estimates**

Design-storm benchmarks were obtained from the NOAA Atlas 14 precipitation frequency dataset, which provides long-term estimates of rainfall intensity and duration across the United States[59]. The 100-year, 24-hour rainfall depth was selected to represent the regional extreme precipitation standard for each study area. Atlas 14 values reflect the expected depth of rainfall with a 1% annual exceedance probability, offering a climatological benchmark against which observed rainfall intensity can be evaluated. The dataset, available at an approximate 800-meter spatial resolution, was converted from inches to millimeters and spatially resampled to align with the MRMS 1-km grid.

### 3.2.3 Crowdsourced Flood Impact Data

To evaluate the correspondence between modeled hazard levels and real-world disruptions, two independent crowdsourced datasets were used.

- **311 Service Requests**

Municipality-maintained 311 service request databases from Harris County and Dallas County capture citizen-reported flood impacts during each event period. These datasets include geocoded records of street flooding, drainage failures, blocked inlets, and related flood complaints submitted by residents to city response systems[60]. For each flood event, all 311 requests corresponding to the 24-hour accumulation period were extracted and aggregated at the H3 x L10 hexagonal grid. The resulting 311 count per hexagon represents the spatial distribution and intensity of reported flood-related disruptions within each county.

- **Waze traffic incident reports**

Geo-referenced traffic incident reports were obtained from the Waze for Cities Program, which aggregates real-time, user-generated information on roadway conditions[61]. Reports tagged as flooded road, road closure, impassable, or hazard-weather were filtered for each event's 24-hour duration. These Waze data provide high-frequency, crowdsourced observations of mobility disruptions linked to flooding. Similar to the 311 dataset, individual Waze reports were spatially joined to H3 L10 hexagons, and the incident count within each hexagon was used as a proxy for the severity of flood-induced transportation impacts.

**Table 1. Summary of the datasets used in this study**

| Data | Description | Source |
|---|---|---|
| Fathom Pluvial Flood Depths | Modeled pluvial flood depths for the 1-in-100-year return period | Texas Water Development Board |
| Distance to Stream$_0$ | Euclidean distance (m) to any stream segment (Strahler $\geq$ 0), capturing local hydrologic connectivity | USGS National Hydrography Dataset |
| Distance to Stream$_4$ | Euclidean distance (m) to major streams (Strahler $\geq$ 4), capturing proximity to trunk river systems | USGS National Hydrography Dataset |
| HAND | Vertical elevation difference between ground surface and nearest drainage cell | University of Texas National Flood Interoperability Experiment |
| MRMS 24-h Accumulated Rainfall | Radar-based quantitative precipitation estimates for 24-h event periods. | NOAA National Severe Storms Laboratory |
| NOAA Atlas 14 | 100-year, 24-hour design-storm rainfall depths | NOAA Hydrometeorological Design Studies Center |
| 311 Service Requests | Geocoded citizen reports of street flooding, drainage failures, and flood-related complaints | City of Houston and City of Dallas Open Data Portals |
| Waze Traffic Incident Reports | Geo-referenced user-reported traffic incidents tagged as flooded roads, closures, or weather-related hazards | Waze for Cities Program |

## 4. Methods

Figure 1 shows the overview of this study. The study comprises two primary components: (a) a hazard assessment framework and (b) a validation approach. The hazard assessment framework operates through three integrated layers. First, the IHL component employs pixel-based classification at H3 Level 10 resolution to characterize persistent landscape vulnerability using three key variables: Fathom flood maps, distance-to-stream networks, and HAND values. This classification produces four risk classes (A through D) representing high to low flood susceptibility based on topographic position and hydrologic connectivity, with high-risk zones (Class A) concentrated in low-lying areas near drainage networks and low-risk zones (Class D) on elevated terrain distant from streams. Second, the HSI integrates the persistent IHL classification with event-specific precipitation data by comparing MRMS 24-hour rainfall observations against NOAA Atlas 14 100-year 24-hour storm thresholds. This comparison generates rainfall ratios (observed/threshold) that quantify precipitation extremes relative to climatological design standards. The framework applies these ratios through class-specific thresholds in the LTS layer, which accounts for differential flood response across landscape positions. Areas with inherently high flood susceptibility (IHL Class A) require lower rainfall ratios to trigger elevated threat levels, while inherently low-risk areas (IHL Class D) require substantially higher rainfall ratios to reach equivalent threat designations. This produces a matrix of 20 unique threshold values defining five severity levels (A through E) for each of the four IHL risk classes, enabling spatially heterogeneous hazard characterization that reflects both precipitation intensity and underlying landscape vulnerability. The validation approach employs two independent crowdsourced datasets to evaluate framework performance against observed flood impacts: municipal 311 service requests documenting citizen-reported flooding requiring municipal response, and Waze traffic incident reports capturing real-time traffic disruptions associated with flooding. The study applies this validation methodology to two distinct flood events: Tropical Storm Imelda (Harris County, September 2019) and a flash flood event (Dallas County, August 2022). Both datasets are spatially aggregated to H3 hexagons and normalized using min-max scaling to create a composite impact index through equal-weighted combination (50-50 weighting). Spearman rank correlation analysis quantifies the monotonic relationship between parametric severity classifications (LTS levels) and observed impacts, with statistical significance assessed through p-value testing. This validation design tests whether hexagons classified with higher parametric severity levels exhibit systematically greater observed impacts as measured by crowdsourced data.

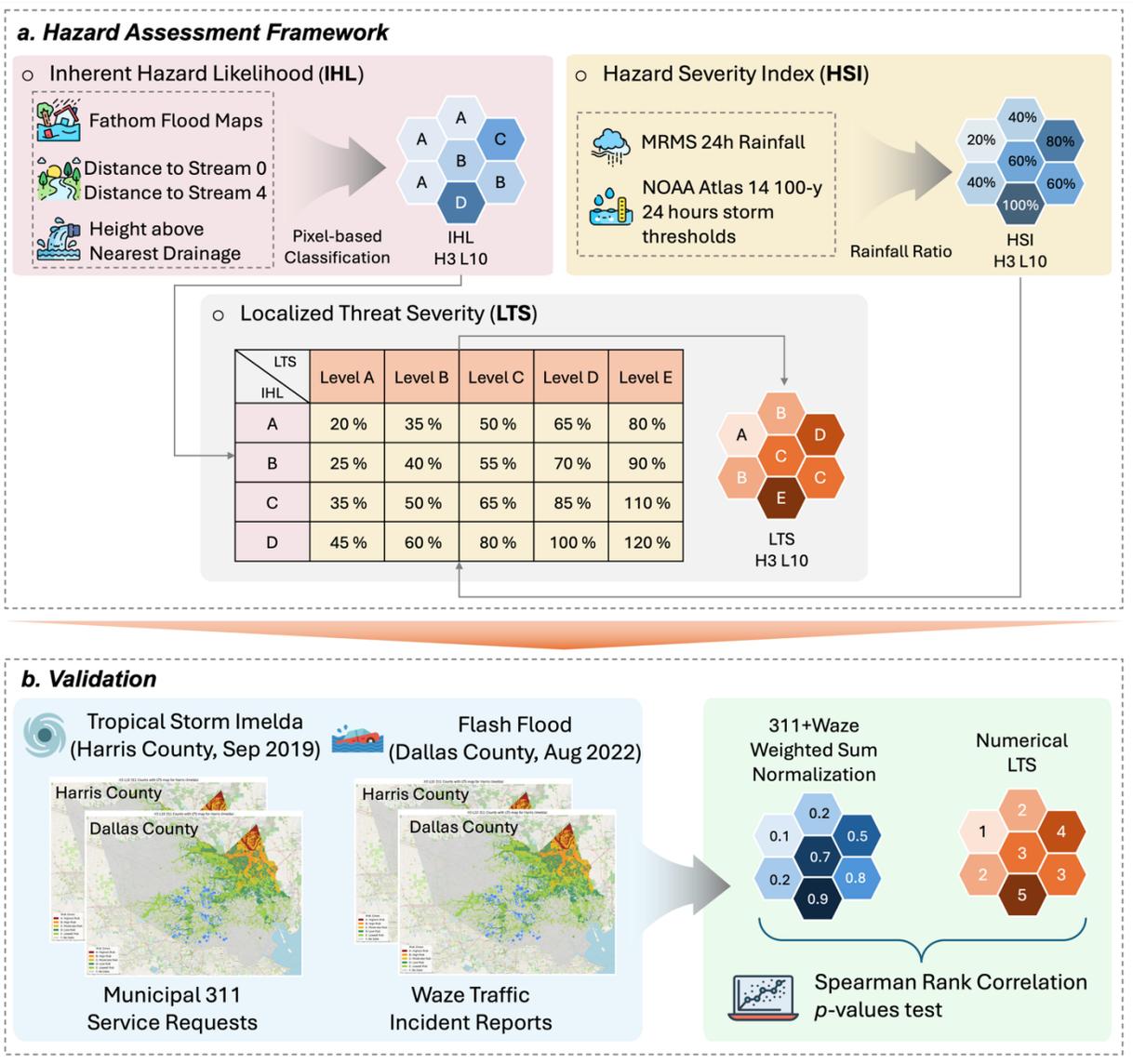

**Figure 1. Overview of the study.** (a) Hazard-assessment pipeline: static landscape susceptibility is mapped as inherent hazard likelihood by fusing Fathom 1-in-100-yr pluvial depth, HAND, and distance-to-stream data; event severity is expressed as a hazard severity index by taking the ratio of MRMS 24-hour rainfall to local NOAA Atlas 14 100-year/24-hour depth; IHL and HSI are combined on an H3 level 10 grid using a class-specific localized threat severity matrix (4 IHL classes × 5 severity levels = 20 thresholds) to produce neighborhood/watershed-scale warning maps. (b) Validation pipeline: independent crowdsourced impact data (311 service requests and Waze traffic incidents) are aggregated to H3, min-max normalized, and combined into an equal-weighted composite impact index; Spearman rank correlation tests whether higher localized threat severity levels align with greater observed disruption. The framework is demonstrated on two contrasting Texas events (Tropical Storm Imelda, Harris County, 2019; Dallas flash flood, August 2022) to assess transferability across hydrologic regimes

- **Inherent hazard likelihood (IHL)**

This study implemented a pixel-based classification scheme to delineate flood risk into four hierarchical zones through the integration of Fathom flood maps, dual distance-to-stream datasets, and height above nearest drainage variables. To ensure spatial consistency, all input rasters were resampled and aligned to a common grid framework, with the HAND layer serving as the reference coordinate system. Table 2 shows the criteria by which we classify risks.

**Table 2. IHL risk category and criteria**

| Risk Category | Criteria |
| --- | --- |
| A. High Risk | Fathom $\geq$ 1.64 or<br>HAND $\leq$ 2 and Distance to Stream0 < 100 |
| B. Moderate Risk | 0.5 $\leq$ Fathom < 1.64 or<br>2 2< HAND $\leq$ 5 and Distance to Stream4 $\leq$ 300 |
| C. Urban / Shallow Flood | HAND $\leq$ 3 and Fathom > 0 |
| D. Low / No Risk | Else condition |

The IHL taxonomy is intentionally grounded in physically interpretable predictors—modeled pluvial depth (Fathom), vertical position relative to drainage (HAND), and hydrologic connectivity (distance to streams)—ensuring that each class captures a distinct, well-understood flood mechanism. Class A (high risk) identifies pixels with either deep modeled pluvial flooding (Fathom $\geq$ 1.64 m), or low-lying, channel-proximate terrain (HAND $\leq$ 2 m within 100m of any stream), representing areas primed for rapid overbank or channel-adjacent inundation. Class B (moderate risk) encompasses pixels with substantial but sub-critical modeled depths (0.5–1.64 m) or slightly higher terrain (HAND 2–5 m) close to major channels (Distance to Stream$_4$ $\leq$ 300m), reflecting common floodplain widths and backwater or overbank spread along trunk systems. Class C (urban/shallow flood) isolates shallow, low-lying settings (HAND $\leq$ 3 m) and Fathom is larger than 0. All other pixels default to Class D (low/no risk). These thresholds leverage Fathom's LISFLOOD-FP foundations to represent depth-driven hazard, while HAND and stream proximity encode topography and connectivity—variables repeatedly demonstrated to be among the strongest predictors of flood susceptibility.

To facilitate multi-scale analysis and integration with other geospatial datasets, the pixel-based risk classifications were aggregated to the H3 hierarchical hexagonal grid system at resolution level 10. Within each hexagon, the majority risk class among all contributing pixels determined the hexagon's final classification, effectively summarizing fine-resolution risk patterns at a standardized coarser spatial unit.

- **Hazard Severity Index (HSI)**

This study utilized high-resolution radar-derived quantitative precipitation estimates from the Multi-Radar Multi-Sensor (MRMS) system, which provided gridded 24-hour accumulated rainfall at approximately 1-km spatial resolution. Rainfall values were extracted and standardized to millimeters to ensure measurement consistency across datasets. For comparative analysis, design storm rainfall depths were derived from NOAA Atlas 14 100-year 24-hour precipitation frequency estimates for Texas. Atlas 14 rainfall depths, originally provided in thousandths of an inch at approximately 800-meter resolution, were converted to millimeters for consistency with observed data. Both MRMS and Atlas 14 datasets were subsequently resampled and spatially aligned with the H3 level 10 hexagonal grid framework established during the risk classification phase to facilitate direct spatial comparison and analysis. These paired metrics enabled direct comparison between actual event precipitation and statistically derived design storm values at a consistent spatial resolution. A dimensionless rainfall ratio was calculated for each hexagon by dividing the MRMS by the Atlas 14 100-year 24-hour design depth. This metric quantifies the severity of the observed event relative to the local design storm standard. Values exceeding 1.0 indicate rainfall depths that surpassed the 100-year return period threshold, while values below 1.0 represent events of lesser magnitude than the design standard.

- **Localized threat severity (LTS)**

To integrate rainfall severity with the pre-existing flood risk classification, this study developed a tiered hazard assessment framework employing class-specific rainfall ratio thresholds. Lower ratio thresholds were established to trigger elevated hazard levels in Class A zones. Conversely, low-risk areas (Class D), situated on elevated terrain distant from streams, demonstrate greater resilience to extreme precipitation and therefore require higher ratio thresholds to reach equivalent hazard designations. Table 3 establishes class-specific rainfall-ratio triggers (HSI = 24-hour MRMS divided by local Atlas-14 100-year depth) to convert the same forecast or observed precipitation into landscape-aware alert levels. Thresholds are intentionally lowest in Class A (10–70%) because areas with low HAND values and stream-proximate terrain flood with comparatively modest rainfall exceedance, while Class D requires substantially larger ratios (25–120%) to avoid false alarms on uplands with weak hydrologic connectivity. The step sizes follow a non-linear progression, with smaller increments at Levels A through C and larger increments at Levels D through E, reflecting the physics of flash flooding where impacts accelerate rapidly once drainage capacity is exceeded. This design also provides operational stability by reducing oscillations near high-consequence triggers. Expressing triggers as percentages of the local design storm maintains comparability across Texas's diverse rainfall gradient while ensuring the system remains both auditable and intuitive for emergency managers and insurers. For instance, Level E in Class C clearly indicates that rainfall has reached or exceeded 100% of the 100-year 24-hour depth at that specific pixel. The 20-cell matrix structure, comprising four classes and five levels, provides sufficient granularity to accommodate heterogeneous urban and fluvial settings while remaining sufficiently compact for effective calibration and prospective tuning as additional events and ground truth data become available.

**Table 3. Localized threat severity criteria**

| HSI \ LTS | Level A | Level B | Level C | Level D | Level E |
|---|---|---|---|---|---|
| **Class A** | 10% | 20% | 30% | 50% | 70% |
| **Class B** | 15% | 25% | 35% | 55% | 90% |
| **Class C** | 20% | 30% | 40% | 60% | 100% |
| **Class D** | 25% | 35% | 45% | 65% | 120% |

- **Flood risk model verification**

To validate the parametric hazard severity classifications against observed flood impacts, this study integrated two independent crowdsourced datasets: Waze traffic incident reports and municipal 311 service requests from Hurricane Imelda (2019) and the 2022 August Dallas flood event. These datasets provide complementary perspectives on flood-related disruptions, with Waze capturing real-time traffic impediments and 311 requests documenting citizen-reported flooding requiring municipal response.

The Waze dataset comprised georeferenced traffic reports submitted by application users during flood events, filtered to include only flood-related incident categories, such as road closures, hazardous conditions, and traffic jams associated with flooding. The 311-dataset included citizen service requests related to street flooding, drainage issues, and flood-related infrastructure failures, geocoded to street addresses. Both datasets were spatially joined to the H3 level 10 hexagonal grid, with incident counts aggregated within each hexagon to produce hexagon-level impact metrics.

To create a unified measure of observed flood impact, this study normalized and integrated the Waze and 311 count variables. Both count variables were normalized using min-max scaling to transform raw counts to a unitless scale ranging from 0 to 1. This transformation standardized the two variables to comparable scales while preserving relative magnitude relationships within each dataset. In cases where all hexagons exhibited identical counts (zero variance), normalized values were set to zero to avoid undefined division operations. An equal-weighted composite impact index was then constructed by computing the arithmetic mean of the normalized Waze and 311 variables for each hexagon. This 50-50 weighting scheme reflects the assumption that traffic disruptions and municipal service requests provide equally valid but independent indicators of flood severity. The resulting composite index ranged from 0 to 1, with higher values indicating greater observed flood impact.

For validation analysis, only hexagons meeting three criteria were retained: (1) non-missing values for both Waze and 311 counts, and (2) hazard severity classifications other than "F" (no data). This filtering criterion isolated hexagons experiencing both measurable rainfall exceedance and observed impacts, providing a robust basis for validating the hazard severity framework. The relationship between parametric hazard severity classifications and observed flood impacts was evaluated using Spearman rank correlation analysis. The Spearman correlation coefficient ($\rho$) quantifies the monotonic association between two ordinal or continuous variables without assuming linearity or normal distributions, making it appropriate for relating categorical hazard severity levels (A through E) to continuous impact indices. For each hexagon in the filtered dataset, the hazard severity level (expressed as an ordinal variable) was paired with its corresponding composite impact index value. The Spearman rank correlation coefficient and associated p-value were computed to assess the strength and statistical significance of the monotonic relationship. A positive correlation would indicate that hexagons classified with higher hazard severity levels tend to exhibit greater observed impacts, thereby validating the parametric framework's predictive capability. The p-value tests the null hypothesis of no monotonic association, with values below 0.05 indicating statistically significant correlation at the 95% confidence level.

## 5. Results

Figure 2 presents the spatial distribution of H3 level 10 hexagons classified by inherent flood hazard likelihood across Harris County, Texas. The map depicts four distinct risk zones: high-risk areas (Class A, red) concentrated along major stream corridors and downstream reaches; moderate-risk zones (Class B, orange) in transitional floodplain areas; urban/shallow flood-prone areas (Class C, yellow) distributed throughout developed regions; and low-risk locations (Class D, green) occupying elevated terrain distant from drainage networks. The classification reveals a dendritic pattern of high-risk hexagons aligned with the natural drainage network, including the San Jacinto River system in the eastern portion of the county and Buffalo Bayou traversing the urban core. Moderate and low-risk zones form a heterogeneous matrix across interfluves and upland areas, reflecting variations in topographic position and hydrologic connectivity. This spatial framework provides the inherent flood hazard level.

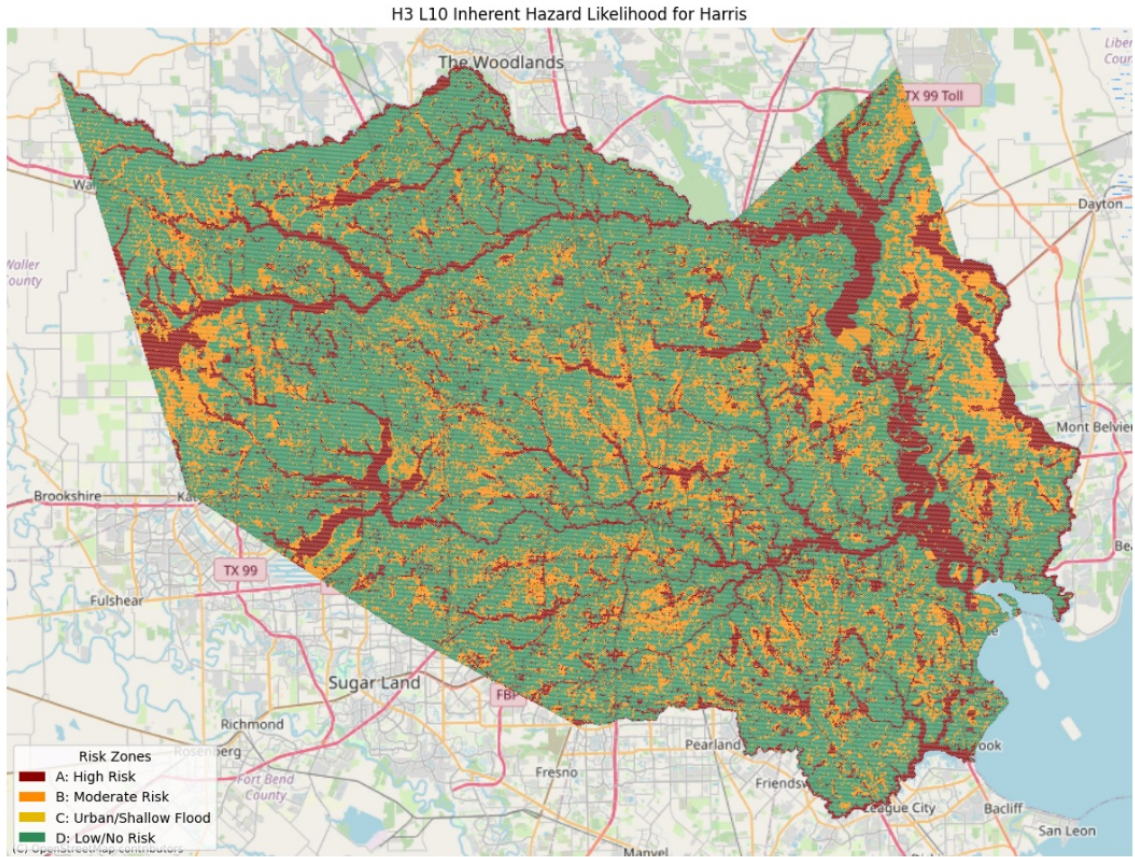

**Figure 2. Harris County inherent hazard likelihood.**

Figure 3 presents the results of hazard severity assessment and crowdsourced impact validation for Tropical Storm Imelda between 11:30 p.m., September 17, 2019, and 11:00 p.m., September 18, 2019, in Harris County. Panel (a) displays the H3 L10 hazard severity index, showing a continuous gradient of severity values with the most extreme conditions (yellow-green) concentrated in areas in the eastern portion of the county. Most areas in the Harris County exhibit moderate severity values (purple-blue), indicating relatively uniform but less extreme rainfall exceedance. Panel (b) illustrates the localized threshold severity classification, revealing discrete risk categories with high-risk zones (Level A, red) and moderate-risk zones (Level B, orange) in the eastern and northeastern sections of the county. Large areas of western Harris County display gray shading (Class F, no data), indicating regions where observed rainfall remained below design storm thresholds. Panels (c) and (d) overlay crowdsourced impact data on the localized threat severity map to validate the parametric classification framework. Panel (c) shows the spatial distribution of 311 service requests, with blue points representing citizen-reported flooding incidents concentrated in the southern and central portions of the county, broadly aligning with areas classified as moderate to high risk. Panel (d) displays Waze traffic incident reports, also represented as blue points, showing a similar spatial pattern with concentrations in central Harris County.

The validation analysis yielded a statistically significant and positive correlation between the localized threshold severity classifications and the combined crowdsourced impact index for Tropical Storm Imelda. The equal-weighted composite index, integrating normalized 311 service requests and Waze traffic incident reports, produced a Spearman rank correlation coefficient of ρ = 0.004 with a p-value of 0.026. While the p-value below 0.05 indicates statistical significance at the 95% confidence level, the correlation coefficient magnitude of 0.004 suggests a weak monotonic relationship between parametric hazard severity levels and observed flood impacts.

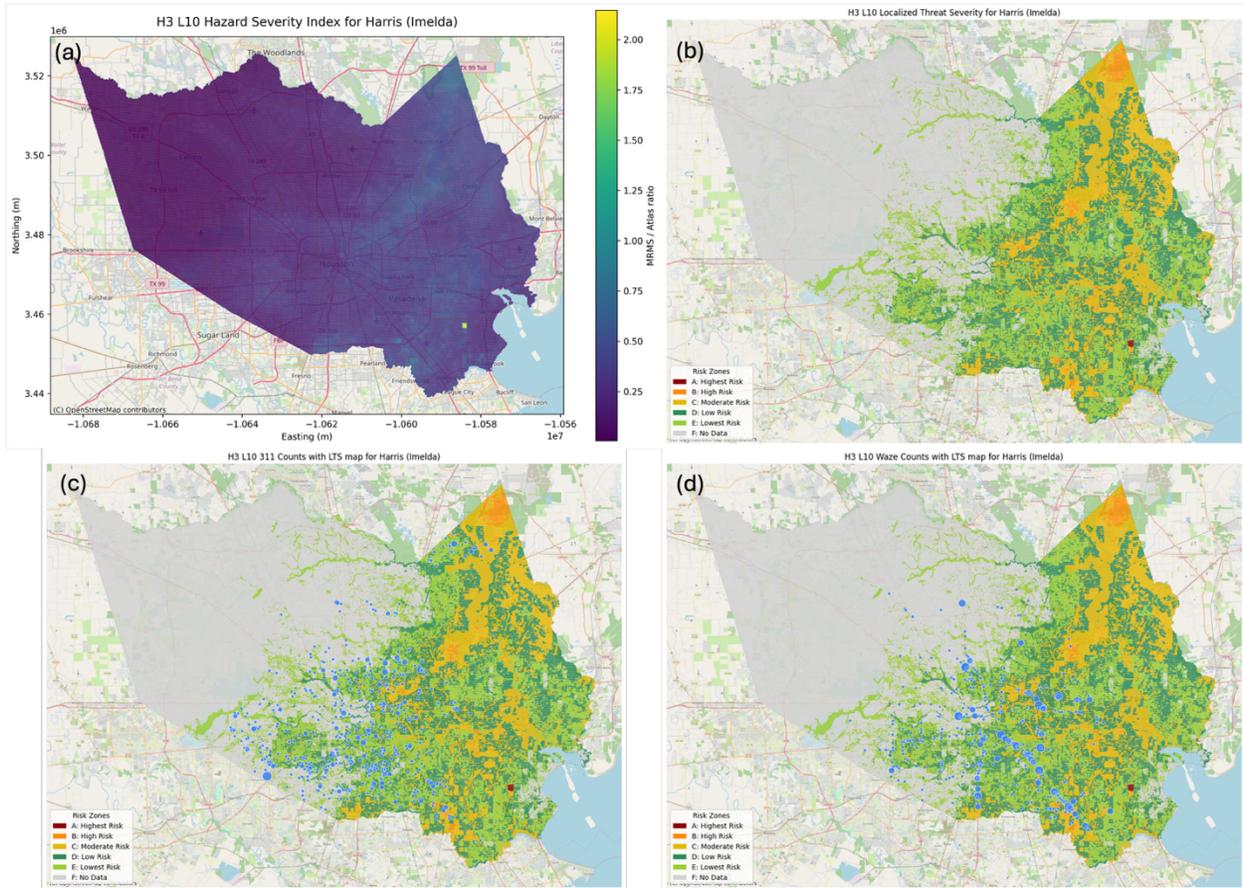

Figure 3. Results of Tropical Storm Imelda between 11:30 p.m., September 17, 2019, and 11:00 p.m., September 18, 2019. (a). H3 L10 hazard severity index; (b). H3L10 localized threshold (LTS) severity; (c). H3 L10 311 counts with LTS map; (d). H3 L10 Waze counts with localized threshold map.

Figure 4 presents the hazard severity assessment and crowdsourced impact validation for Tropical Storm Imelda during the 24-hour period from 11:00 p.m., September 18, 2019, to 11:00 p.m., September 19, 2019, in Harris County. Panel (a) displays the H3 L10 hazard severity index, showing a northeast-to-southwest gradient with the highest severity values (yellow-green) concentrated in the northeastern quadrant of the county. This spatial pattern contrasts markedly with the previous 24-hour period (Figure 2), indicating the temporal evolution of extreme rainfall during Imelda's passage. The majority areas of Harris County exhibits moderate to high severity values (blue-green), suggesting more widespread rainfall exceedance compared to the September 17–18 period. Panel (b) illustrates the localized threshold severity classification, revealing extensive areas classified as high-risk (Level A, red) and moderate-risk (Level B, orange) concentrated in the northeastern portion of Harris County, with additional moderate-risk zones (Class C, yellow) distributed throughout central and eastern areas. The white areas in western Harris County represent regions where rainfall remained below design storm thresholds (Level F). This expanded spatial extent of elevated severity classifications reflects the broader geographic impact during this 24-hour period. Panels (c) and (d) overlay crowdsourced impact data on the localized threat severity map. Panel (c) shows 311 service requests as blue points, densely concentrated in central and southern Harris County, with notable clustering in areas classified as moderate to high risk. Panel (d) displays Waze traffic incident reports, exhibiting extensive distribution across the county with particularly high concentrations in central urban areas and along major transportation corridors. The substantial increase in both 311 and Waze reports compared to the previous 24-hour period (Figure 2) reflects the peak impact phase of Tropical Storm Imelda.

The validation analysis for this time yielded a Spearman rank correlation coefficient of $\rho = 0.113$ with a p-value of 0.000764 between the localized threshold severity classifications and the equal-weighted composite impact index. While the correlation coefficient remains relatively weak in magnitude, the extremely low p-value indicates highly significant statistical association. This improved correlation compared to the September 17-18 period ($\rho = 0.116$, p=5.12e-20) suggests that the parametric hazard severity framework demonstrates greater predictive validity during peak rainfall conditions. The stronger correspondence between severity classifications and observed impacts during this period reflects both the broader spatial extent of extreme rainfall and the concentration of crowdsourced reports in areas experiencing the most severe conditions, as evidenced by the spatial alignment visible in panels (c) and (d).

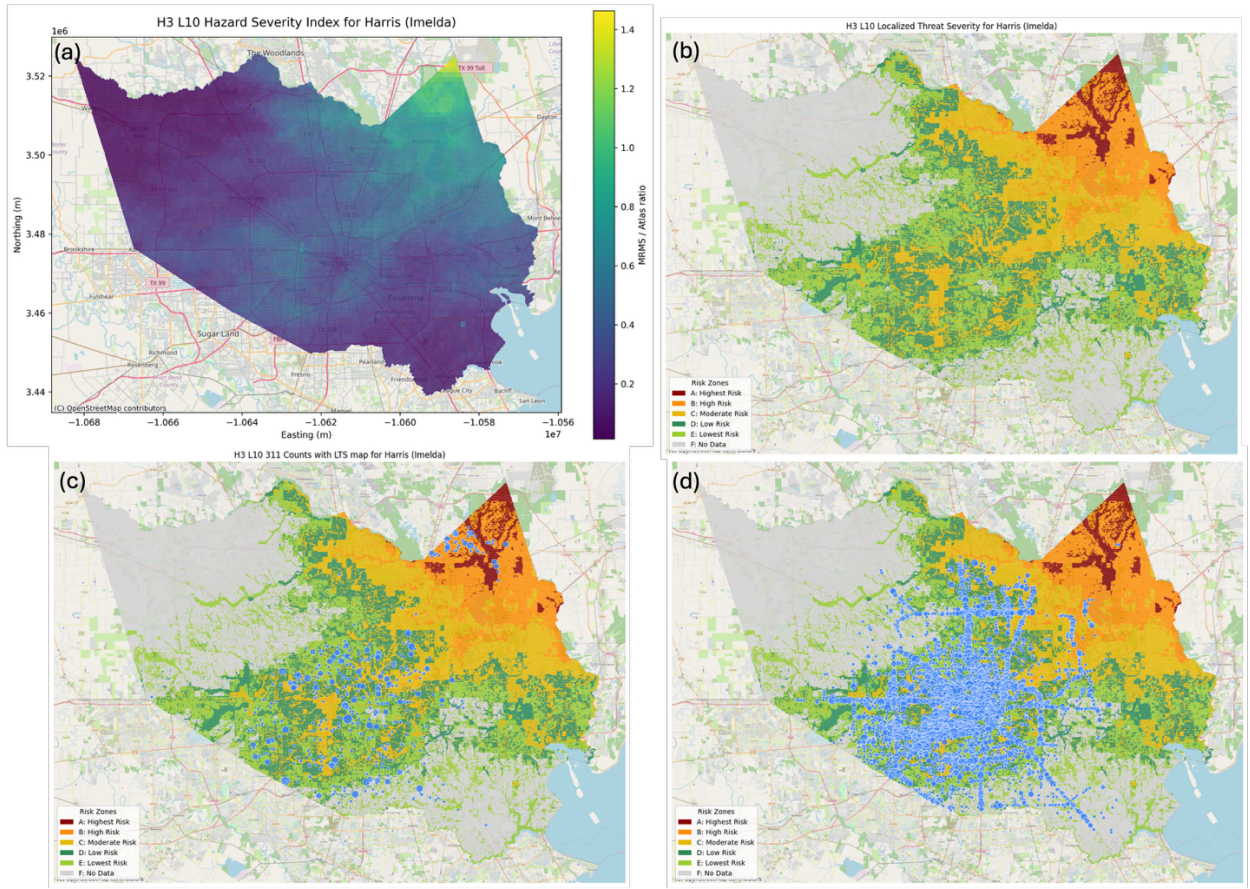

**Figure 4. Results of Tropical Storm Imelda in Harris County in the 24-hour period between 11:00 p.m., September 18, 2019, and 11:00 p.m., September 19, 2029.** (a). H3 L10 Hazard Severity Index of Imelda in Harris County; (b). H3L10 localized threshold severity; (c). H3L10 311 counts with LTS map; (d). H3 L10 Waze counts with LTS map.

Figure 5 presents the spatial distribution of inherent hazard likelihood across Dallas County, Texas, in the 24-hour period from 11 p.m., August 21, 2022, 23:00 through, 11:00 p.m., August 22, 2022, classified into four hierarchical risk zones using the H3 level 10 hexagonal grid framework. The map reveals high-risk areas (Class A, dark red) concentrated along major stream corridors, most prominently following the Trinity River system that traverses the county from northwest to southeast through the downtown Dallas core. These high-risk hexagons form distinct linear patterns aligned with the drainage network, reflecting low topographic positions and direct hydrologic connectivity to channel systems. Moderate-risk zones (Class B, orange) occupy transitional floodplain areas adjacent to primary watercourses and appear as buffer zones between high-risk corridors and upland terrain, particularly evident in the southeastern quadrant where the Trinity River floodplain broadens.

Urban/shallow flood-prone areas (Class C, yellow) are scattered throughout the county, predominantly in developed regions where impervious surfaces and altered drainage patterns create localized flooding potential independent of proximity to major streams. Low-risk areas (Class D, green) dominate the spatial distribution, comprising the majority of Dallas County's land area and representing elevated terrain with minimal hydrologic connectivity to stream networks. This classification pattern demonstrates the strong topographic and hydrologic controls on flood susceptibility, with the dendritic structure of high and moderate-risk zones mirroring the natural drainage network geometry. The framework provides a baseline hazard assessment applicable to rainfall-based severity analysis for future flood events in Dallas County.

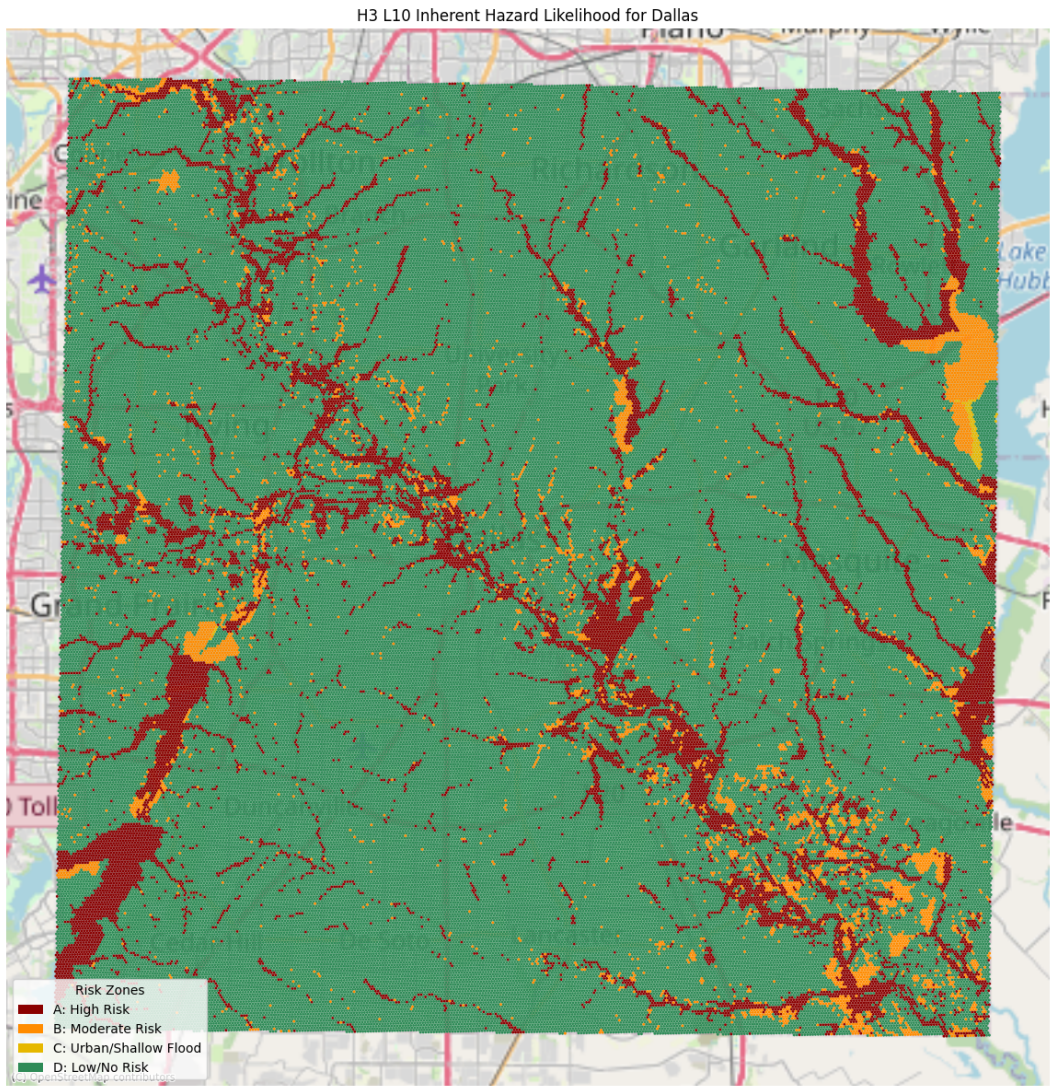

**Figure 5. Dallas County inherent hazard likelihood**

Figure 6 presents the hazard severity assessment and crowdsourced impact validation for the August 2022 flash flood event in Dallas County during the 24-hour period from 11:00 p.m., August 21, 2022, through 11:00 p.m., August 22, 2022. Panel (a) displays the H3 L10 hazard severity index, revealing a pronounced east-west gradient with the highest severity values (yellow, 1.2-1.4) concentrated in the eastern and southeastern portions of the county. Multiple distinct hotspots of extreme severity appear in the eastern quadrant, indicating localized areas where observed rainfall substantially exceeded design storm thresholds. The central and western portions of the county exhibit moderate severity values (blue-green, 0.4–0.8), while northwestern areas show the lowest severity indices (dark blue, 0.2–0.4), reflecting the spatial heterogeneity of rainfall during this convective event. Panel (b) illustrates the localized threshold severity classification, showing a clear dichotomy between eastern and western Dallas County. High-risk zones (level A, dark red) and moderate-risk zones (Level B, orange) dominate the eastern half of the county, forming extensive contiguous areas along the Trinity River corridor and its tributaries. The classification reveals particularly severe conditions in the southeastern quadrant, where multiple high-risk hexagons cluster around major drainage features. Low-risk areas (Level C, yellow-green) and areas below threshold exceedance (Level E, light green; and Level F, white) predominate in western Dallas County, indicating that extreme rainfall was primarily confined to the eastern portion of the study area during this event. Panels (c) and (d) overlay crowdsourced impact data on the localized threat severity map to validate the classification framework. Panel (c) shows 311 service requests as blue points, demonstrating substantial concentration in central and eastern Dallas County with notable clustering in areas classified as high to moderate risk. The spatial distribution of 311 reports exhibits strong correspondence with elevated severity zones, particularly along the Trinity River corridor through downtown Dallas. Panel (d) displays Waze traffic incident reports, revealing extensive distribution of blue points across the county with the highest densities in central urban areas and the eastern quadrant. The Waze data show broader geographic coverage than 311 reports, reflecting real-time traffic disruptions across the regional transportation network.

The validation analysis yielded a Spearman rank correlation coefficient of $\rho = 0.120$ with a p-value of 0.00412 between the localized threshold severity classifications and the equal-weighted composite impact index combining normalized 311 and Waze data. While the correlation coefficient magnitude remains relatively modest, the extremely low p-value indicates highly significant statistical association between parametric severity levels and observed flood impacts. This result demonstrates that the hazard classification framework successfully captures spatial patterns of flooding consequences during the August 2022 event. The visual alignment between crowdsourced impact locations and elevated hazard severity classifications, particularly in the eastern half of Dallas County, provides complementary qualitative support for the parametric framework's predictive capability in identifying areas experiencing actual flood impacts during this convective rainfall event.

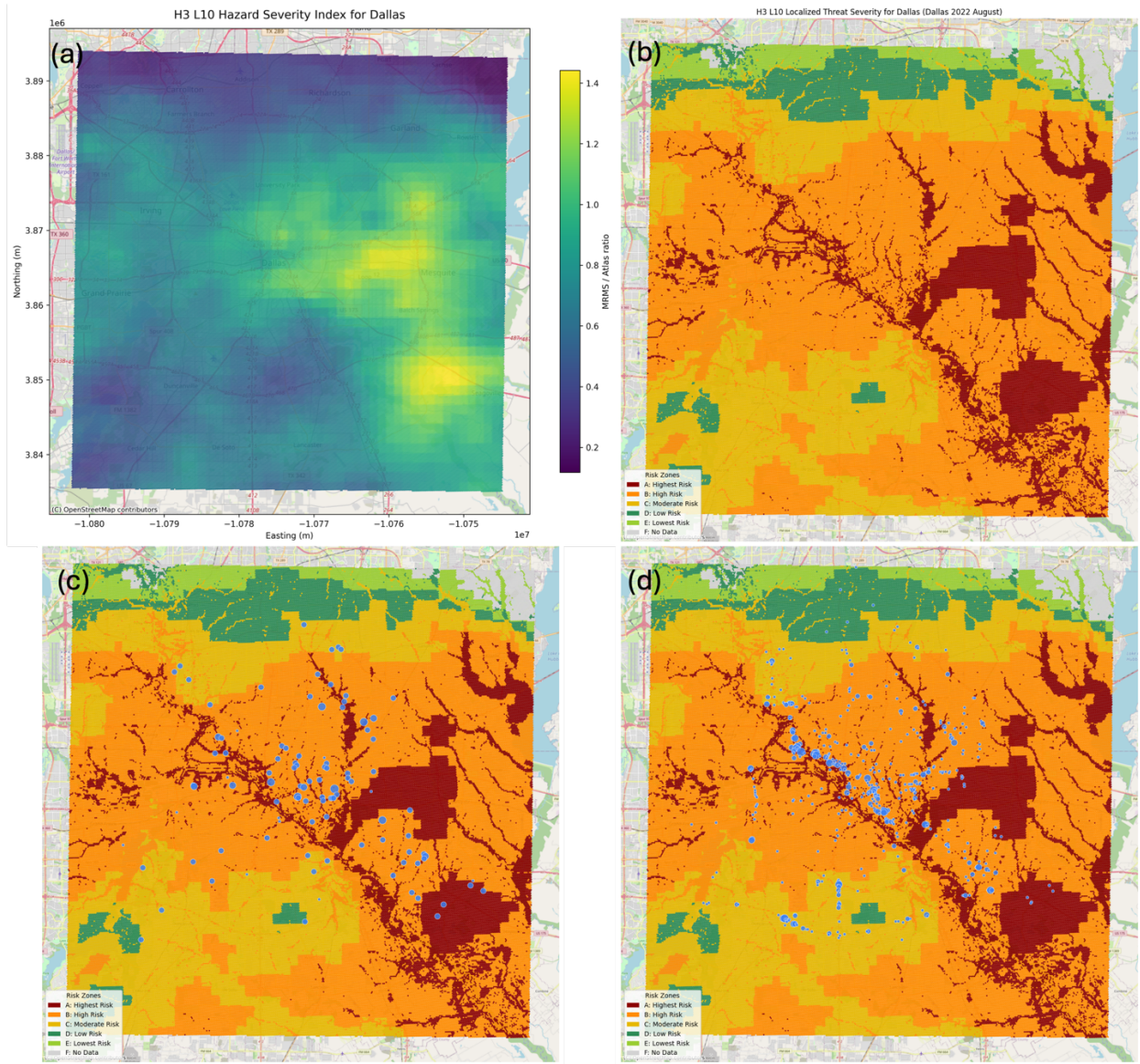

**Figure 6. Results of Dallas flood event in August 2022 event in the 24-hour period between 11:00 p.m., August 21, 2022, and 11:00 p.m., August 22, 2022.** (a). H3 L10 hazard severity index; (b). H3L10 localized threshold severity; (c). H3 L10 311 counts with LTS map; (d). H3L10 Waze counts with LTS

## 6. Discussion and Concluding Remarks

This study introduces an anticipatory, parametric architecture that converts landscape susceptibility and design-storm-normalized rainfall into transparent, zone-aware thresholds at fine spatial resolution. Flash flood warning systems have traditionally relied on real-time sensor data and hydrologic–hydraulic modeling to provide near-term alerts, typically delivering warnings hours before flooding occurs. While these systems have proven highly effective for immediate response, they offer limited lead time for anticipatory actions such as evacuation planning, resource prepositioning, and community preparation. Many proactive flood risk reduction strategies—including large-scale evacuations, prepositioning of emergency resources, activation of emergency operations centers, public preparation messaging, and coordination across multiple jurisdictions—require longer lead times than traditional warning systems can provide. Furthermore, the implementation of detailed flood models requires substantial computational resources, technical expertise, and comprehensive data infrastructure—barriers that many resource-constrained communities face.

To address this critical temporal gap, this study developed and validated a spatially explicit parametric flood hazard assessment framework that enables anticipatory flood warning 48 hours ahead of extreme precipitation events, identifying high-risk areas before flooding occurs. While current systems effectively use real-time sensor data and flood models for near real-time warnings, this parametric approach extends the temporal horizon of flood risk assessment from hours to days before events, when quantitative precipitation forecasts become available but before hydrologic conditions develop sufficiently for real-time modeling. Importantly, this framework is not intended to replace existing real-time flood warning systems, but rather to complement the current suite of flood warning solutions by addressing the anticipatory phase of emergency management that traditional sensor-based and model-driven systems cannot serve. The methodology integrates static topographic and hydrologic variables with dynamic precipitation data to characterize flood risk across multiple temporal and spatial scales. Through application to Hurricane Imelda (September 2019) in Harris County and the August 2022 flood event in Dallas County, the framework demonstrated robust capacity to delineate spatial patterns of flood risk using the H3 hexagonal grid system. The hazard susceptibility index, derived from distance to streams, height above nearest drainage values, Fathom flood maps, and stream network characteristics, successfully identified topographic and hydrologic controls on flood susceptibility. High-risk zones aligned closely with major drainage corridors and low-lying floodplain areas, while low-risk areas occupied elevated interfluves distant from stream networks. This baseline classification provides a persistent representation of landscape vulnerability that remains valid across diverse meteorological conditions, establishing a foundation for event-specific severity assessments. Integration of Multi-Radar/Multi-Sensor observed rainfall with NOAA Atlas 14 design storm thresholds enabled quantitative evaluation of precipitation extremes during actual flood events. The localized threat severity framework applies class-specific rainfall ratio thresholds to account for differential flood response across risk zones, producing spatially heterogeneous severity classifications reflecting both rainfall intensity and landscape position. Results from historical events revealed pronounced spatial gradients in hazard severity corresponding to storm precipitation distributions, with the most extreme conditions concentrated in northeastern Harris County during period of September 18 through19, 2019, while the August 2022 Dallas event exhibited distinct east-west severity gradients aligned with convective rainfall patterns. Validation against crowdsourced 311 service requests and Waze traffic incident reports yielded statistically significant correlations between parametric severity classifications and observed impacts across all validation cases, with highly significant p-values ($p < 0.05$ to $p < 10^{-25}$) confirming systematic relationships between framework predictions and ground-truth observations. While correlation magnitudes remained modest ($\rho$ = 0.004 to 0.020), reflecting confounding factors inherent in crowdsourced data including spatial biases related to population density and reporting behavior, the consistent statistical significance across diverse events and study areas supports the methodology's validity for operational applications.

This framework's scalability and accessibility represent significant advantages for communities lacking resources for detailed flood modeling or advanced technical expertise. The parametric approach leverages static topographic datasets available nationally and operational precipitation products that are freely accessible, requiring neither specialized modeling expertise nor high-performance computing infrastructure. This democratizes access to anticipatory flood hazard information across diverse jurisdictions and socioeconomic contexts, extending capabilities to resource-constrained communities that cannot maintain sophisticated real-time modeling systems. The complementary relationship between parametric anticipatory assessment and real-time warning systems creates a layered flood risk management approach spanning multiple temporal scales. In the 48- to 72-hour window before an event, the parametric framework identifies areas of elevated risk based on forecasted precipitation and landscape vulnerability, enabling proactive preparations, including resource prepositioning, public messaging, and inter-agency coordination. As the event approaches and real-time hydrologic data become available in the 6- to 12- hour window before flooding, traditional warning systems provide precise, localized flood predictions that guide immediate tactical response actions. This temporal stratification ensures that communities benefit from both strategic anticipatory planning and tactical near-term warnings. The framework's broader applicability extends beyond the subtropical coastal and continental interior settings examined in this study. The methodology's reliance on nationally available datasets and standardized precipitation frequency estimates enables transfer to other geographic contexts, climatic regions, and jurisdictional scales. The hexagonal tessellation approach offers computational advantages over traditional administrative boundaries while maintaining consistent spatial resolution, facilitating regional- to continental-scale implementations. By demonstrating a practical methodology for translating meteorological forecasts into spatially distributed flood hazard assessments that integrate landscape vulnerability with precipitation extremes, this research establishes a foundation for anticipatory flood risk management systems that can inform emergency response, public warning, and risk communication during extreme precipitation events. The approach proposed extends actionable lead time to the 48- to 72-hour anticipatory window using open national datasets and provides an operational bridge between static susceptibility maps and reactive hydrologic-hydraulic warnings. Emergency managers can leverage these validated methods to enhance real-time decision-making during flood events, from prioritizing resource deployment and road closures to identifying vulnerable communities requiring immediate assistance. The methodology's robustness across multiple study areas further supports its transferability to other cities and regions, enabling flood management agencies to integrate crowdsourced intelligence into existing decision support systems without requiring extensive local calibration. Understanding the confounding factors identified in this analysis, such as spatial autocorrelation and varying reporting rates, allows practitioners to interpret crowdsourced signals appropriately while maintaining confidence in their operational utility for improving flood resilience and emergency response outcomes.

The present study is intentionally lightweight and anticipatory, but some constraints merit attention. First, event severity is expressed as a 24-hour MRMS-to-Atlas 14 ratio, which inherits Atlas 14's stationarity assumptions and may underrepresent short-duration bursts (1 to 3 hours) that often drive flash flooding. Second, validation relies on crowdsourced proxies (311 and Waze) that are informative but biased by population density and reporting behavior, which helps explain the modest yet significant correlations. Third, classification at H3 L10 can aggregate fine-scale urban drainage controls that influence local flooding patterns. Fourth, thresholds were set a priori rather than statistically calibrated across regions or prospectively tested with quantitative precipitation forecasts. Future research should pursue several enhancements: (i) adopt multi-duration triggers (15-minute, 1-hour, 3-hour, 6-hour) and incorporate forecast ensembles to evaluate lead-time skill; (ii) integrate antecedent conditions (soil moisture, tide and reservoir stage) and urban drainage attributes (storm-sewer capacity, pump stations) to modulate local thresholds; (iii) expand ground truth to include DOT road-closure logs, National Flood Insurance Program claims, high-water marks, and damage surveys for stronger, less-biased validation; (iv) downscale to multi-scale H3 (L10 to L12) for dense urban cores where finer resolution is critical; (v) perform hierarchical or Bayesian calibration of class-specific thresholds across hydrologic regions with reliability diagrams and Brier decompositions; (vi) run prospective real-time pilots to quantify operational latency, false-alarm costs, and decision value; and (vii) benchmark against two-dimensional hydrologic–hydraulic models to identify where parametric triggers add value or require refinement—collectively strengthening fidelity while preserving the framework's computational efficiency and deplorability.

## Data availability

All data used in this study are publicly available and have been properly cited in the paper.

## Code availability

All analyses were conducted using Python. The code that supports the findings of this study is available from the corresponding author upon request.

## Acknowledgements

This work was supported by the National Science Foundation under Grant CMMI-1846069 (CAREER). Any opinions, findings, conclusions, or recommendations expressed in this research are those of the authors and do not necessarily reflect the view of the funding agency.

## Competing interests

The authors declare no competing interests.